\documentclass[12pt]{emulateapj}

\usepackage{graphicx, perpage}
\MakePerPage{footnote}

\shortauthors{Pineda, Bottom, \& Johnson}

 \newcommand{\ks}{K_{s}}
 \newcommand{\mk}{\mathcal{M}_{Ks}}
  \newcommand{\mall}{\mathcal{M}_{JHKs}}
 \newcommand{\vk}{V-\ks}
 \newcommand{\dvk}{\Delta (\vk)}

\begin{document}

\title{Using High-Resolution Optical Spectra to Measure Intrinsic Properties of Low-Mass Stars: New Properties for KOI-314 and GJ$\,$3470}

\author{J. Sebastian Pineda \altaffilmark{1,2} Michael Bottom \altaffilmark{2} and John. A. Johnson\altaffilmark{2}}
\affiliation{California Institute of Technology, Department of Astronomy, 1200 E. California Ave, Pasadena CA, 91125, USA}

\altaffiltext{1}{Corresponding author: jspineda@astro.caltech.edu}
\altaffiltext{2}{California Institute of Technology, Department of Astronomy, 1200 E. California Ave, Pasadena CA, 91125, USA}

\begin{abstract}
We construct high signal-to-noise ``template'' spectra by co-adding hundreds of spectra of nearby dwarfs spanning K7 to M4,  taken with Keck/HIRES as part of the California Planet Search. We identify several spectral regions in the visible (370 - 800 nm) that are sensitive to the stellar luminosity and metallicity. We use these regions to develop a spectral calibration method to measure the mass, metallicity, and distance of low-mass stars, without the requirement of geometric parallaxes. Testing our method on a sample of nearby M dwarfs we show that we can reproduce stellar masses to about 8 - 10\%, metallicity to $\sim 0.15$ dex and distance to 11\%. We were able to make use of HIRES spectra obtained as part of the radial velocity monitoring of the star KOI-314 to derive a new mass estimate of $0.57 \pm 0.05$  $M_{\odot}$, a radius of $0.54\pm0.05$ $R_{\odot}$, a metallicity, [Fe/H], of $-0.28 \pm 0.10$ and a distance of $66.5\pm7.3$ pc. Using HARPS archival data and combining our spectral method with constraints from transit observations, we are also able to derive the stellar properties of GJ$\,$3470, a transiting planet hosting M dwarf.  We estimate a mass of $0.53 \pm 0.05$  $M_{\odot}$, a radius of $0.50\pm0.05$ $R_{\odot}$, a metallicity, [Fe/H], of $0.12 \pm 0.12$ and a distance of $29.9\pm_{3.4}^{3.7}$ pc. 
\end{abstract}

\section{Introduction}

M-type dwarf stars are poorly understood compared to higher-mass FGK stars because of the difficulty in modeling both their atmospheres and interior structures (\citealt{Hauschildt1999, Chabrier2000} ). The cool atmospheres of these stars contain many molecular species such as VO and TiO, which dominate both the line and continuum opacity in their photospheres (\citealt{Kirkpatrick1991}). Due to the millions of molecular transitions, many of the opacity sources have yet to be accounted for, making it difficult to synthesize M-dwarf spectra. Modeling has also proven challenging in this mass regime because convection plays an important role in the structure of the star (\citealt{Baraffe1998, ribas2006, morales2010}). Much of the input physics for low-mass stars remains approximate, and consequently, the physical properties of M dwarfs outside of binary systems are difficult to measure (for example studies of binaries see \citealt{lopezmorales2005, coughlin2011}).

Despite these challenges, there have been many attempts to discern the intrinsic properties of isolated M dwarfs. These efforts have focused on using observational properties to either aid the theoretical treatments or to develop empirical calibrations based on stars of known physical properties. \cite{Delfosse2000} demonstrated a tight relation between absolute magnitudes in the infrared passbands ($J$, $H$, and $K_{s}$) and stellar mass as determined from binary systems. As the metallicity of these stars increases, their colors redden and the bolometric luminosity decreases. These effects balance in the infrared to produce relatively tight mass-luminosity relations. 
 
\cite{Delfosse2000} also suggested that the large scatter in the color-magnitude diagram, ($\vk$) -- $\mk$, was due to this same effect. \cite{Bonfils2005} took advantage of this sensitivity to metallicity to develop a photometric calibration based on wide binaries with an M-dwarf secondary and an FGK primary of known [Fe/H]. The \cite{Bonfils2005} photometric calibration has since been revised using improved V-band photometry (\citealt{JA09}, \citealt{neves2012}) and by using a physical model based guide (\citealt{SS10}). \cite{SS10} and \cite{neves2012} parameterize the metallicity as a function of $\dvk$, which measures how much a given star deviates from the main sequence. These efforts represent significant progress in determining the intrinsic properties of M dwarfs; however they require precise photometry and a parallax measurement. Although these requirements are met for nearby M dwarfs (d $\lesssim$ 15 pc), the lack of parallaxes for fainter, more distant stars severely limits our knowledge of M dwarfs beyond the Solar neighborhood. 
 
 Additionally, spectral synthesis has been used to measure M-dwarf metallicities (\citealt{Woolf2005}; \citealt{Bean2006}; \citealt{onehag2012}). For higher-mass stars, spectral line comparisons based on equivalent width measurements can be used to compute abundances. However, the line blanketing in low-mass stars makes this extremely difficult, as individual lines blend and completely dominate any thermal continuum . \cite{Bean2006} applied atmospheric models to binary systems composed of an FGK primary and an M-dwarf secondary. Assuming a co-evolutionary system, the metallicity of the primary can be determined accurately and used to calibrate the modeling of M-dwarf spectra. However, \cite{Bean2006} were off by  $\sim$ 0.3 dex compared to later photometric calibrations, likely due to deficiencies in the low-mass atmospheric models. \cite{Woolf2005} were successful, but only because they used extremely metal-poor M dwarfs with minimal molecular line blanketing. More recently, \cite{onehag2012} matched synthetic spectra to their high-resolution $J$-band spectra to measure metallicities to within 0.09 dex of the best photometric calibrations.
 
Instead of relying on atmospheric models, recent, more accurate methods have made use of observation-based calibrations. \cite{Rojas2010} used moderate-resolution infrared spectra to develop metallicity indicators based on Ca I and Na I features in the $K$-band. Similarly, \cite{terrien2012} used Ca I and K I features in the $H$-band to measure metallicity. These near-IR spectroscopic calibrations agree well with current photometric calibrations (\citealt{rojas2012}; \citealt{terrien2012}).

Accurately measuring the properties of low-mass stars has gained renewed urgency because of the discovery of a multitude of planets around M dwarfs (\citealt{Butler2004}; \citealt{Rivera2005}; \citealt{BonfilsPlanet}). There are also many M dwarfs among the hosts of planet candidates discovered by the \emph{Kepler} Mission (\citealt{Borucki2011}; \citealt{Kepler2} ; \citealt{muirheadkoi}). Additionally, these stars may host a multitude of terrestrial planets (\citealt{koi961, buchhave2012, swift13}). For radial velocity and transit detected planets, determining the physical properties of the planets requires an accurate measurement of the physical properties of their host stars.
 
 In this contribution we develop a method to measure the physical properties of low-mass stars using spectroscopic indices from high-resolution optical spectra. Specifically, our technique provides estimates of the absolute NIR magnitudes ($\mathcal{M}_J$, $\mathcal{M}_H$, $\mathcal{M}_K$), distances $d$, and $\dvk$ for M dwarfs without parallaxes. Using known observational calibrations these quantities can be converted to estimate both mass and metallicity (\citealt{Delfosse2000, SS10, neves2012}). We create a library of high signal-to-noise, high-resolution ``template'' low-mass dwarf spectra with known photometric properties to develop calibration curves based on the strength of various absorption features. By measuring the strength of these features in the spectrum of a star with unknown properties and using the calibration curves, the photometric properties of the low-mass star can be estimated which in turn can be used to estimate the star's physical properties. In Section~2, we describe the data sample used to develop our library. In Section~3, we cover our analysis methods to construct and develop our calibration curves. In Section~4, we apply our methods to a distant M-dwarf listed among the host stars containing \emph{Kepler} exoplanet candidates and another star with a recently discovered transiting neptune-mass planet. Lastly, in Section~5, we discuss the utility of our methods.

\section{Data} \label{sec:data}

\begin{figure}[htbp] 
   \centering
   \includegraphics[ scale=.5, angle=0]{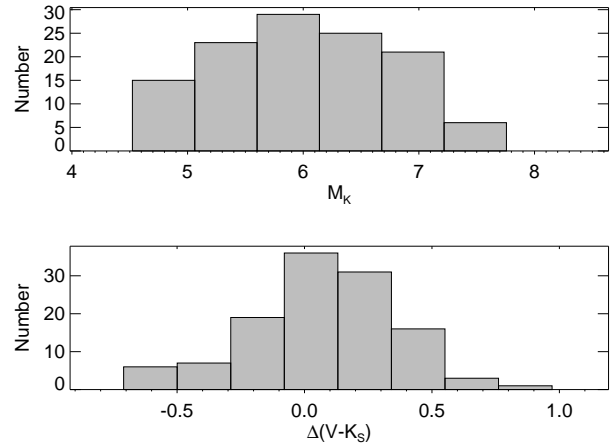} 
   \caption{Distribution of stellar properties in sample. Top panel is $\mk$, and bottom is $\dvk$. The calibration sample thus spans a broad range or stellar properties}
   \label{hist}
\end{figure}

\subsection{Sample}

Over the past decade the California Planet Survey (CPS) has obtained spectroscopic measurements of more than 2,500 stars at Keck Observatory, monitoring their radial velocities for the characteristic signature on the host star induced by the presence of a planet (\citealt{Howard2010}).  We make use of their High-Resolution Echelle Spectrometer (HIRES ; \citealt{vogt1994} ) observations of 155 M dwarfs to build a high-resolution, high signal-to-noise ``template'' spectrum of each star. The typical CPS program HIRES set-up gives a resolving power $R = \lambda / \Delta \lambda \approx 50,000$ and uses an iodine-cell for precise wavelength calibration of  radial velocity measurements (\citealt{Howard2010}). In our application the iodine lines are contaminants, so we can only make use of the blue and red chips of the HIRES detector, where there are no lines from the iodine cell. The blue portion of the spectrum allows us to examine between 370 nm  and 480 nm, while the red portion of the spectrum covers between 650 nm  and 800 nm.

The M dwarfs in the CPS sample have well defined 2MASS photometry and parallaxes from \emph{Hipparcos} (\citealt{Perryman1997}; \citealt{2MASS}). This allows us to characterize the stellar sample in terms of their absolute near-IR magnitudes, ($\mall$) and color, ($\vk$). We culled stars from the sample that were overly active or young, selected on the basis of having published rotation periods less than 5 days, high X-ray luminosities with a X-ray count rate $>1$ count $s^{-1}$ in ROSAT \citep{rosat} or as being members of a young open cluster or moving group. We also culled stars that were known to be unresolved binaries or turned out to be missing geometric parallax measurements. We also limited the sample to stars brighter than $\mk = 8$, so that the range of properties is well sampled by the CPS stars. This left us with 119 calibration stars. The sample spans a range of $\mk$ from $4.5$ to $7.5$ as shown in the top panel of Figure \ref{hist}. Using the \cite{JA09} solar-metallicity main sequence relation, defined as $\mk = \sum a_{i} (\vk)^{i}$, where $a = \{-9.58933, 17.3952, -8.88365, 2.22598, -0.258854,$ $ 0.0113399\}$, we can calculate the quantity, $\dvk$, the difference between the observed color and the main-sequence color for a star of the same $\mk$ such that positive values of $\dvk$ correspond to redder objects. We will henceforth refer to $\dvk$ as the color offset, which can be used as a proxy for metallicity (\citealt{SS10}). The bottom panel of Figure \ref{hist} shows that our sample spans a broad range in color offset and hence a broad range in metallicity.

\subsection{Spectral Library}\label{speclib}

Since our sample of calibration stars have been monitored for radial velocity shifts indicative of planets over the past 10-15 years, each star has an average of 25 observations. By combining the individual observations we can produce high signal-to-noise, high-resolution template spectra for each of the calibration stars in our sample. We rebinned each star's spectrum onto a wavelength scale that is evenly spaced in $\ln{\lambda}$ (cadence of $1.9\times10^{-6}$),  which allows us to properly Doppler shift the spectra with respect to one another (\citealt{Tonry1979}). We also corrected for small differences in the wavelength solution from night to night (on the order of a couple of pixels) for every spectral order due to changes between the cross-disperser angle and the echelle angle as well as changes in the slit illumination for any given observation.

\begin{figure}[htbp] 
   \centering
   \includegraphics[scale=.45,angle=270]{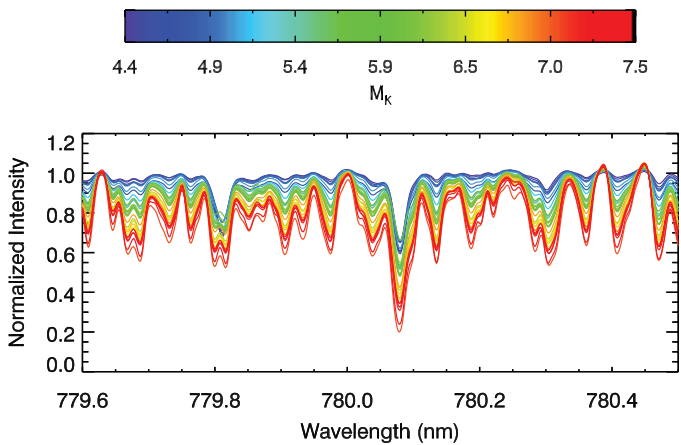} 
   \caption{ Normalized Intensity as a function of wavelength for sensitive regions with bin-averaged spectra of width 0.1 mags in $\mk$. Colors correspond to $\mk$ in the range 4.5 to 7.5, with red being fainter stars and blue being brighter. }
   \label{sens}
\end{figure}

Having aligned the spectra and removed defects (cosmic ray hits etc.), we simply co-added the flux to produce a high signal-to-noise ratio spectrum for each star. The red portions of the spectrum yield a typical signal-to-noise ratio (SNR) per resolution element ranging from $\sim800$ to $\sim4000$, depending on the number of observations for the particular star. For the blue-chip spectra the SNR ranges from $\sim100$ up to $\sim600$. The SNR of the red side is higher primarily because the peak of the M dwarfs' spectral energy distribution lies closer to near-infrared wavelengths than to the blue portion of the spectrum. 

In order to compare the spectra it was necessary to normalize each order to remove the effect of the blaze function of the spectrometer and to account for a pseudo-continuum. We normalize each spectral order individually. First, we separate the order into tens of ``chunks", masking out telluric regions, and take the top 1-2\% of points of each chunk as representing the ``continuum". We then fit a low-order polynomial to the continuum points across the full extent of the order. Lastly, we divide the spectral order by this polynomial, ignoring problematic points at the ends of the orders, to get the normalized spectrum (see Appendix for further details). Since the echelle spectra are not flux calibrated and the blaze function distorts the shape of the spectra, a pseudo-continuum is not well defined through our normalization procedure. However, because we use the same continuum regions for all stars, this process allows for reliable, differential comparisons among stars of different types.

In Figure \ref{sens}, we plot the template spectra averaged in $\mk$ bins with width of 0.1 mag. The different colors correspond to different values of $\mk$ with red corresponding to cooler stars and blue to hotter stars. In the Figure, we see that the absorption features are quite distinct from the ``continuum" , which match across the different stars. It is clear in the high signal-to-noise spectra that features such as these are quite sensitive to $\mk$ and we can use the strength of the absorption for a given spectrum as indicative of the stars intrinsic luminosity. Deviations from a strict sequence are primarily due to metallicity effects. At a given $\mk$ changes in metallicity will affect the strength of certain absorption features. Accounting for this second-order effect  will provide us with a valuable indicator of stellar metallicity.

\section{Analysis}

\subsection{Spectral Calibration}

The presence of spectral regions sensitive to changes in $\mall$ motivates the development of a quantitative relationship between the strength of each region and the physical properties of a star. The regions that were found to be sensitive and useful for calibration are listed in Table \ref{spectable}. We identified these regions by eye looking across the full spectrum, ignoring telluric regions, and requiring the continuous regions of the spectrum to have have monotonically increasing absorption with decreasing stellar effective temperatures (see Figure \ref{sens}). The useful regions consist predominately of portions of TiO and VO bands. Note that the spectral indices listed in the table are all in the red portion of the spectrum. Although there also appeared to be sensitive regions at blue wavelengths, the lower overall flux level limits their usefulness, both for the calibration procedure and for future observations. In addition to the regions listed in Table \ref{spectable} there were other regions that we identified as sensitive to the physical stellar properties. However, we selected a subset that when combined provided the optimal calibration (see Section~\ref{sec:acc}).

The CPS sample spans a representative range of properties for low-mass stars, making it useful for our calibration procedure. Since the sample is limited in size and individual stars only represent discrete points in the mass-metallicity plane, comparing the spectra directly is less than ideal, giving poor parameter resolution. Instead, it is preferable to compare the strength of sensitive features, measured from their integrated fluxes (equivalent widths, EWs), allowing us to fit smooth functions to observed trends in EW and providing a continuous relationship between the integrated flux and the stellar properties. The integrated flux is defined as

\begin{deluxetable}{c c c}
\tablecaption{ Spectral Indices \label{spectable}}
\tablehead{ \colhead{ Center } &  \colhead{ Total}  & \colhead{ Molecule / } \\ 
 \colhead{ Wavelength (nm) } &  \colhead{ Width (nm)}  & \colhead{  Line  \tablenotemark{a}}  }
\startdata
 658.24 & 0.820 & TiO   \\
 660.35 & 0.59 & TiO   \\
 663.89 & 0.81 & TiO   \\
 668.82 & 1.71 & TiO   \\
 710.27 & 2.91 & TiO   \\
 713.64 & 2.48 & TiO \tablenotemark{b}  \\
 727.01 & 0.06 & TiO   \\
 770.04 &  0.79 & TiO / KI   \\
 770.83 & 0.50 & TiO  \\
 776.92 & 2.38 & TiO \\
 780.13 & 0.92 & TiO \\
 787.30 & 0.85 & TiO / VO  \\
 791.04 & 0.51 & VO \\
 792.26 & 1.67 & VO  \\
 793.26 & 0.30 & VO \\
 794.87 & 0.24 & VO 
\enddata
\tablenotetext{a}{Predominant Molecules and Lines based on spectra of \citealt{Kirkpatrick1991}}
\tablenotetext{b}{This region is closely related to the TiO4/5 bands defined in \citealt{Reid1995}}
\end{deluxetable}

\begin{eqnarray}
\mathrm{EW} &=& \Delta \lambda - \int S(\lambda) d\lambda  \\
 &\approx& \Delta \lambda - \sum_{i} S_{i} h_{\lambda} \; ,
 \label{eq:approx}
\end{eqnarray}

\noindent where  $\Delta \lambda$ is the width over which the integral is computed and $S(\lambda)$ is the normalized spectrum as a function of wavelength $\lambda$. Equation \ref{eq:approx} gives the approximation for discretely sampled spectra over pixels that span $\Delta\lambda$ evenly sampled with width $h_{\lambda}$. To determine the errors on our equivalent width measurements we randomly simulate the spectral observation using poisson statistics with a mean in each spectral bin given by the photon counts of the actual data. We take the error on the equivalent width measurement to be the standard deviation of the distribution of EW values in the simulation. The typical error on the EW measurements are $\sim 1-5\%$.

\begin{figure}[htbp] 
   \centering
   \includegraphics[width=.45\textwidth,angle=0]{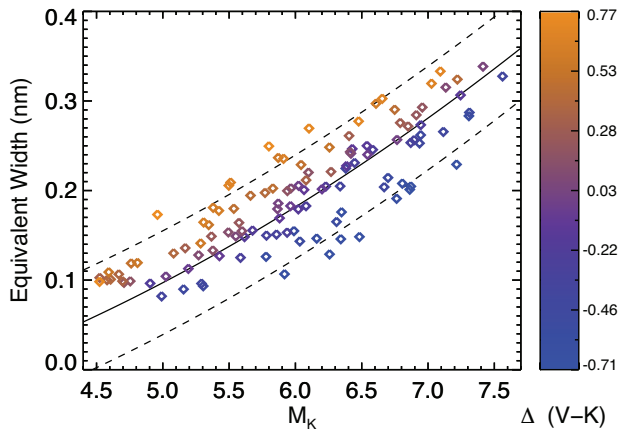} 
   \caption{ EW plotted as a function of $\mk$ for all stars in the sample. The colors are ordered according to the $\dvk$ of the stars in the sample. There is a clear gradient that corresponds to differences in metallicity. The lines correspond to contours in our polynomial model of constant color offset, $\dvk$, with values 0.5, 0.0 and -0.5 going from top to bottom respectively. This region is centered at  770.04 nm  with a width of 0.50 nm}
   \label{metal_index}
\end{figure}

In Figure \ref{metal_index} we plot the behavior of EW as a function of $\mk$ and $\dvk$ for a particular spectral region. The colors correspond to $\dvk$ with red corresponding to stars that are more metal-rich and blue corresponding to stars that are more metal-poor. For spectral regions such as this one, the strength of the feature increases with increasing metal content as well as decreasing luminosity. The behavior, expressed in Figure \ref{metal_index}, motivates the parameterization of each EW in terms of $\mall$ and $\dvk$. For a given region, $l$, we calibrate the equivalent widths against each of the absolute magnitudes and the color offset. Our calibration EW, which allows us to interpolate between the discrete sample star properties, is given by

\begin{equation}
\mathrm{EW}_{l, \alpha} = ( b +  c \mathcal{M}_{\alpha} ) \dvk + \sum^2_{i=0} a_{i} \mathcal{M}_{\alpha}^{i} \; .
\label{eq:ewtheory}
\end{equation}

\noindent where $\alpha \in \{ \; J ,\; H, \; Ks \; \}$, so there is a separate calibration for each passband using the same spectral region. 

We fit each passband separately instead of  going directly to mass, despite tight mass-luminosity relations, because the infrared colors (ex. $J-K$) are not simply functions of mass and can change with metallicity; this is in contrast to other broadband photometric studies of low-mass stars (\citealt{johnlhs, john254} ). In addition to the simple polynomial terms expressed in Equation \ref{eq:ewtheory}, we include a cross term governed by the coefficient $c$ that accounts for differences between brighter and fainter stars in how their absorption strength responds to the addition of metals. The need for such a term is evident in how TiO features are known to saturate in late-type M dwarfs and that VO features are not apparent in early M dwarfs but appear in late-type M dwarfs (\citealt{Kirkpatrick1991}). 

\subsection{Fitting Broadband Photometry}\label{sec:fit}

We used a Bayesian method to fit Equation \ref{eq:ewtheory} to each set of EWs. In addition to the coefficients in Equation \ref{eq:ewtheory}, we also incorporated an additional parameter, $\sigma$, to take into account intrinsic scatter in our choice of parameterization. Using a Markov Chain Monte Carlo (MCMC) technique with a Metropolis-Hastings algorithm, we explore the posterior distribution for the best parameters in the calibration conditioned on the known properties of the stars in the sample. The best calibration parameters are those that maximize their respective marginal distributions, and thus maximize the probability of each parameter reproducing the stellar properties of the stars in the sample with our simple model. We report the parameter values of our calibration in Table \ref{tab:coeffs}. For a given spectral index the first row of the table corresponds to the parameters for the calibration with $\mathcal{M}_{J}$, the second row for  $\mathcal{M}_{H}$ and the third for $\mk$.

\begin{deluxetable*}{c c c c c c c}
\tablecaption{ Calibration Parameters \tablenotemark{a} \label{tab:coeffs}}
\tablehead{ \colhead{ Center (nm) } &  \colhead{ $a_{0}$ (nm)} &  \colhead{ $a_{1}$ (nm)} &  \colhead{ $a_{2}$ (nm)} & \colhead{$b$ (nm) } & \colhead{$c$ (nm) } & \colhead{$\sigma$ (nm) }  }
\startdata
658.24 &-0.4501058 &0.09198486 &-0.001427114 &0.07882096 &0.002933661 &0.01270252 \\
  &-0.3490226 &0.07719585 &-0.0005361921 &0.06294231 &0.006100980 &0.01208604 \\
  &-0.3486802 &0.07969309 &-0.0005093491 &0.07282996 &0.005513314 &0.01123767 \\ [4pt]
660.35 &-0.1596480 &0.01916057 &0.001997996 &0.001896713 &0.008210727 &0.007939490 \\
  &-0.1520269 &0.02420063 &0.001671400 &0.003454885 &0.009034893 &0.007594586 \\
  &-0.1553699 &0.02603111 &0.001767160 &0.01073864 &0.008722165 &0.007079715 \\[4pt]
663.89 &-0.4226041 &0.08017676 &-0.00008847487 &0.07432834 &0.003435445 &0.01253907 \\
  &-0.3302306 &0.06765508 &0.0007076964 &0.05919351 &0.006587219 &0.01168440 \\
  &-0.3803160 &0.08669426 &-0.0005393973 &0.08166828 &0.004118372 &0.01082510 \\[4pt]
668.82 &-2.836934 &0.7375158 &-0.03934088 &0.6871808 &-0.06801754 &0.03053437 \\
  &-2.196094 &0.6257768 &-0.03455122 &0.5937715 &-0.05868804 &0.02922317 \\
  &-2.374936 &0.7089150 &-0.04205847 &0.6733981 &-0.07173435 &0.02711429 \\[4pt]
710.27 &-5.145841 &1.292010 &-0.06165890 &1.147429 &-0.08640261 &0.06357273 \\
  &-4.044212 &1.116565 &-0.05456582 &1.010953 &-0.07058973 &0.06100981 \\
  &-4.307040 &1.244481 &-0.06565496 &1.162449 &-0.09311087 &0.05549368 \\[4pt]
713.64 &-6.357848 &1.673048 &-0.08776160 &1.399011 &-0.1214399 &0.05989609 \\
  &-5.054271 &1.471289 &-0.08102439 &1.272483 &-0.1101706 &0.05812801 \\
  &-5.253722 &1.593891 &-0.09263240 &1.423004 &-0.1339820 &0.05169340 \\[4pt]
727.01 &-0.1138363 &0.02944591 &-0.001551650 &0.02305361 &-0.001910082 &0.001294007 \\
  &-0.09486709 &0.02712001 &-0.001531907 &0.02163241 &-0.001814339 &0.001242313 \\
  &-0.09856623 &0.02938447 &-0.001748182 &0.02431960 &-0.002235494 &0.001166170 \\[4pt]
770.04 &-0.1598857 &0.008933704 &0.006038753 &0.1399386 &-0.004808356 &0.01121777 \\
  &-0.1181027 &0.008006942 &0.006416439 &0.1134480 &-0.0005393014 &0.01118438 \\
  &-0.1085764 &0.004873543 &0.007252633 &0.1158942 &-0.000001015819 &0.009809055 \\[4pt]
770.83 &-0.3815610 &0.07699559 &-0.001319326 &0.04795545 &0.003966268 &0.009098964 \\
  &-0.2753433 &0.05733777 &-0.00003557394 &0.03629728 &0.006487797 &0.008746648 \\
  &-0.3191640 &0.07383029 &-0.001163489 &0.05463092 &0.004446250 &0.007962008 \\[4pt]
776.92 &-1.408374 &0.2440679 &0.001240992 &0.1658279 &0.02825762 &0.04063114 \\
  &-1.204047 &0.2346570 &0.001405473 &0.1440118 &0.03585729 &0.03961519 \\
  &-1.302366 &0.2751659 &-0.0008869298 &0.2068376 &0.03001058 &0.03584416 \\[4pt]
780.13 &-0.3511850 &0.05775816 &0.001359554 &0.05715288 &0.007504470 &0.01079916 \\
  &-0.2988942 &0.05584809 &0.001437870 &0.05035518 &0.009766074 &0.01034952 \\
  &-0.2972508 &0.05705502 &0.001681733 &0.06221154 &0.009053781 &0.009174584 \\[4pt]
787.30 &-0.07969772 &-0.01437250 &0.005574401 &0.009402555 &0.01204589 &0.01161012 \\
  &-0.04158101 &-0.02027625 &0.006402968 &-0.003833748 &0.01558149 &0.01126873 \\
  &-0.06175869 &-0.01473584 &0.006440116 &0.005074938 &0.01547640 &0.01026771 \\[4pt]
791.04 &0.01844099 &-0.02735590 &0.004700167 &-0.02789386 &0.01186093 &0.006290352 \\
  &0.01743679 &-0.02505636 &0.004849061 &-0.02857221 &0.01330172 &0.006062947 \\
  &-0.0009558277 &-0.02019831 &0.004791414 &-0.02164277 &0.01312965 &0.005528097 \\[4pt]
792.26 &0.2785090 &-0.1369218 &0.01704969 &-0.1025768 &0.03602162 &0.01778133 \\
  &0.2310036 &-0.1224368 &0.01723172 &-0.09838984 &0.03933845 &0.01729166 \\
  &0.2100510 &-0.1211605 &0.01818554 &-0.09387771 &0.04123889 &0.01595478 \\[4pt]
793.26 &0.1454432 &-0.04912226 &0.004548514 &-0.01954720 &0.006100179 &0.003336126 \\
  &0.1161707 &-0.04273733 &0.004429765 &-0.02024410 &0.006868234 &0.003257263 \\
  &0.1180228 &-0.04515902 &0.004855676 &-0.02172495 &0.007529807 &0.003095342 \\[4pt]
794.87 &0.04278556 &-0.02045305 &0.002583565 &-0.004188786 &0.004094394 &0.002292025 \\
  &0.03241636 &-0.01713182 &0.002513652 &-0.002884554 &0.004370116 &0.002306728 \\
  &0.02124669 &-0.01420778 &0.002431369 &-0.0002786377 &0.004293440 &0.002088636 
\enddata
\tablenotetext{a}{The first row for a given index center corresponds to parameters for the $J$ calibration, the second row for $H$ calibration and third row for $K_{s}$ calibration}
\end{deluxetable*}

These curves provide a calibration, which we can apply to stars of unknown properties with measured EWs and estimate $\mall$ and $\dvk$. Taking the errors in the integrated flux as normally distributed, we minimize a $\chi^2$ statistic to get the stellar photometric properties. The statistic is split up into a couple components, 

\begin{equation}
\chi^{2} =  \sum_{l} \chi^{2}_{l} + \chi^{2}_{\mathrm{p}}  \; ,
\label{eq:chitot}
\end{equation}

\noindent where the first term comes from the spectral calibration and is summed over the different indices, $l$, and the last term correspond to photometric constraints. The sum over all the different indices allows us to partially break the degeneracy between metallicity and mass inherent in the individual indices.  

 For each spectral region, $l$, the fitting statistic is 

\begin{eqnarray}
\chi^{2}_{l} &=& \sum_{\alpha} \frac{ [  \mathrm{EW}^{obs}_{l} - \mathrm{EW}_{l,\alpha}(\dvk, \mathcal{M}_{\alpha}) ]^{2}   }{2(\sigma^{2}_{obs,l}  + \sigma_{l,\alpha}^{2})} \; , \end{eqnarray}

\noindent where EW$_{obs, l}$ is the measured EW, with uncertainty $\sigma_{obs,l}$, and $\mathrm{EW}_{l,\alpha}$ is from Equation \ref{eq:ewtheory}, using the best calibration parameters ;  here the scatter parameter, $\sigma_{l,\alpha}$, is added in quadrature to the measurement error. The sum is over each of the three infrared passbands, $\alpha \in \{ \; J ,\; H, \; Ks \; \}$, which have separate calibrations for the given spectral region (see Table \ref{tab:coeffs}). 

Since many nearby stars have 2MASS photometry, we can use the distance, $d$, as an additional parameter by requiring that our estimates of $\mall$ reproduce the observed photometry. 

\begin{equation}
\chi^{2}_{\mathrm{p}} = \sum_{\alpha} \frac{ \{ \alpha - [\mathcal{M}_{\alpha} + 5\log_{10} (d/10 \mbox{pc})  \; ]\}^{2}}{2 \sigma^{2}_{p,\alpha}  } \; ,
\end{equation}

\noindent where $\alpha$ corresponds to the observed infrared magnitudes and $\sigma_{p,\alpha}$ is the corresponding measurement uncertainty. 

Summing over all of the indices and the additional constraints gives the total $\chi^{2}$ of Equation \ref{eq:chitot}, which we minimize as a function of $\mall$, $\dvk$ and $d$ to determine the best-fit stellar properties and the distance to the star.

\begin{figure}[htbp]
   \centering
   \includegraphics[scale=.35,angle=0]{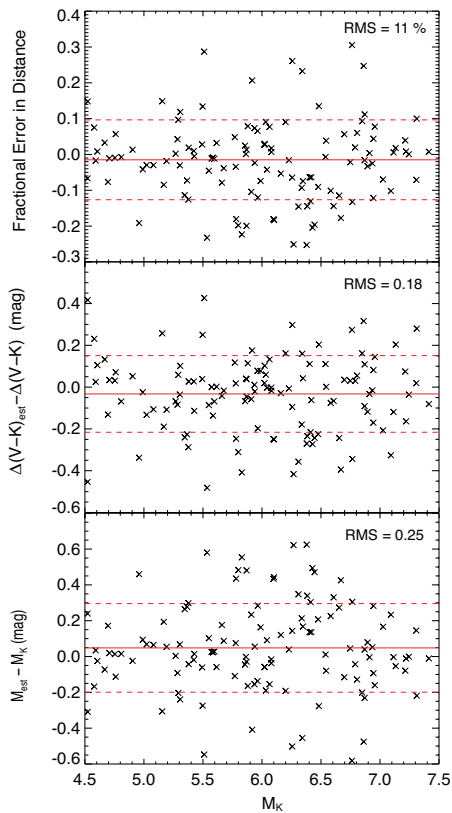} 
   \caption{The error in the distance (top panel), color offset (middle panel) and $\mk$ (bottom panel) as a function of the intrinsic brightness of the star. We can reproduce the distances to an accuracy of 11$\%$, the color offset to 0.18 dex and $\mk$ to 0.25 dex. The RMS in the absolute magnitudes for J and H band closely match the RMS in the K band absolute magnitude.  There is no clear trend, as the method appears to be uniformly applicable between $ 4.5 < \mk < 7.5$. The solid lines mark the mean of the example set and the dashed lines mark the 1$\sigma$ levels about the mean.}
   \label{fig:errors}
\end{figure}

\subsection{Accuracy} \label{sec:acc}

In Figure 4 we show the results of our assessment of how well we can recover the properties of the stars in our calibration sample. The top panel of the Figure shows the percentage error in reproducing the observed distance of the stars, the middle panel shows the error on the color offset and the bottom panel shows the error in $\mk$. The root-mean-scatter (RMS) for the distance, color offset and absolute K-band magnitude are $11\%$, 0.18 dex and 0.25 dex respectively. Using established photometric relations, this scatter would correspond to 0.10-0.15 dex in [Fe/H] (depending on the literature calibration) and $\sim$ 0.05 $M_{\odot}$ in mass (\citealt{SS10} ; \citealt{Delfosse2000}).

\begin{figure}[htbp]
   \centering
   \includegraphics[width = .45\textwidth]{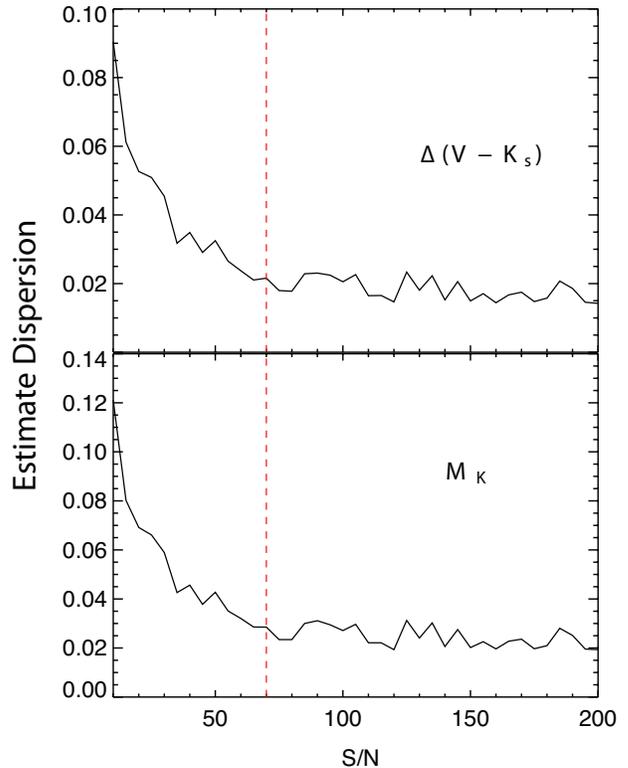} 
   \caption{The degree to which the parameter estimates vary as a function of the median signal-to-noise ratio of the input spectrum measured across the red chip of the HIRES detector. When the input spectrum has a SNR above $\sim$70 (marked by dashed line), the parameter estimates settle down to a well defined value. See text for analysis procedure.}
   \label{fig:snrtest}
\end{figure}

As an additional test, we examined what SNR is necessary to get consistent parameter estimates from any given stellar spectrum using our methods. We simulated a given SNR by adding noise to our template spectra using a pseudo-random number generator. Repeating this many times for each SNR, we saw what effect the noise had on our parameter estimates. In Figure \ref{fig:snrtest}, we plot the dispersion in our estimates as a function of SNR. The top panel is for $\dvk$ and bottom for $\mk$. As the SNR increases, the initial improvements are significant. But after about a SNR of $\sim$ 70, the improvements with greater signal are marginal.

\subsection{Spectral Resolution}\label{sec:res}

\begin{figure}[htbp]
   \centering
   \includegraphics[width=.46\textwidth]{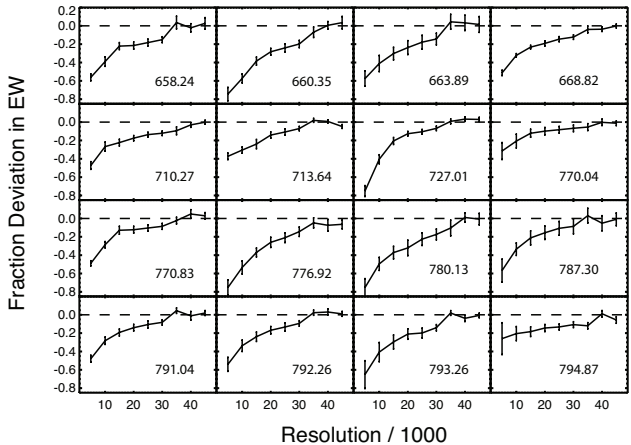}
   \caption{The average fractional deviation in measured EW compared to our original HIRES spectra as a function of spectral resolution for each spectral index designated by the central wavelength (nm) listed in Table \ref{spectable}. Each point represents the mean deviation across the full calibration sample with the error bars given by the corresponding scatter. There are consierable deviations for resolutions below $\sim$30,000.}
   \label{fig:resol}
\end{figure}

We also examined how spectral resolution affects the utility of our calibrations. By measuring EWs consistent with our original measurements we would be able to use our calibration to recover the same set of stellar properties using lower resolution spectra. To test this, we first took the template spectra for our calibration sample and convolved the spectra down to lower resolutions (5,000-45,000) in increments of 5,000 from our original resolution of 50,000. We then normalized the spectra in the same manner as we did our original template spectra (see Section~\ref{speclib}) and computed the EW for each of the indices of Table \ref{spectable} and all of the sample stars. In Figure \ref{fig:resol}, we plot the average fractional difference in EW measurements, across all the calibration stars, between the convolved spectra and the unconvolved spectra as a function of spectral resolution, where each panel corresponds to a different index as listed in Table \ref{spectable}. The error bars in the plot represent the scatter in the deviation across the sample of calibration stars. For each index the EW measurements are consistent with the original measurements for spectral resolutions above 30,000. Therefore our calibration should not be used below this threshold without accounting for the systematic effects demonstrated in Figure \ref{fig:resol}. Although the integrated flux of a spectrum should not change at lower resolutions the blending of pseudo-continuum with the many absorption band-heads complicates our EW measurements.

\begin{figure}[htbp]
   \centering
   \includegraphics[width=.46\textwidth]{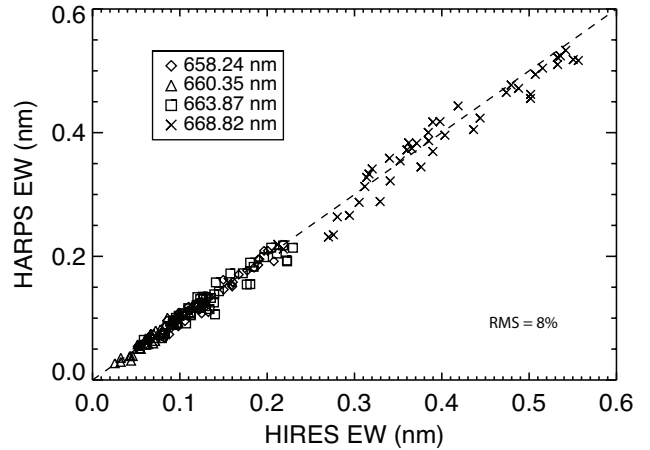}
   \caption{A comparison of the EW measurements for the subsample with HARPS spectra. The different symbols correspond to the four different indices available and the line indicates a 1:1 correspondence. The scatter about the line is at the 8\% level consistent with the measurement uncertainties.}
   \label{fig:harpsew}
\end{figure}

To consider higher resolution data we examined a subsample within our set of calibration stars that also had publicly available data from the ESO data archive using the HARPS spectrograph at a resolution of 115,000 \citep{harps}. For this subsample of 43 stars we combined multiple observations and put together template spectra similar to our HIRES spectra (see Section~\ref{speclib}). This produced a set of HARPS spectra with high SNRs, all greater than 70. We then convolved the spectra down to the HIRES resolution of 50,000 to compare EW measurements. Normalizing the spectra in the usual manner we then measured the EWs for the first four indices of Table~\ref{spectable}; because of the small overlap between the HARPS spectrograph and the HIRES red chip only these four indices were available. In Figure~\ref{fig:harpsew}, we compare equivalent width measurements from the HARPS subsample to the measurements from our HIRES spectral templates. The straight line in the plot corresponds to exact agreement. The measurements agreed rather well with an RMS of $\sim8\%$, in line with the combined errors between the two measurements.

\section{Applications}

\subsection{KOI-314}

\begin{figure}[htbp]
   \centering
   \includegraphics[width = .47\textwidth ]{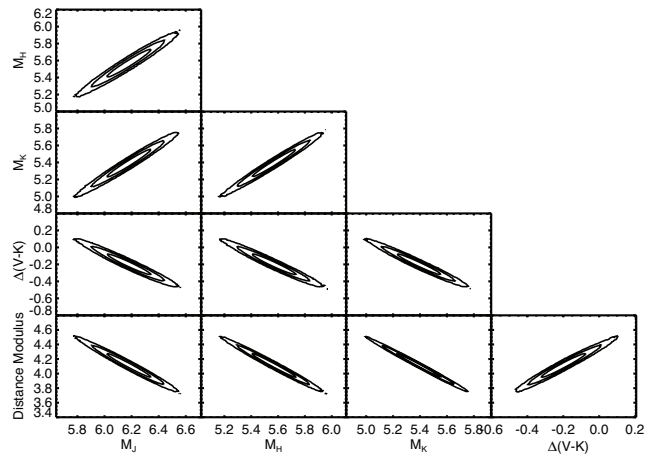} 
      \caption{Contour plots of the posterior probability distribution from the five parameter MCMC analysis of KOI-314. The contour levels represent the $1\sigma$, $2\sigma$, and $3\sigma$ confidence levels. The oblong contours come from spectral indices being degenerate in $\mk$ and $\dvk$}
   \label{fig:contours314}
\end{figure}

We applied our methods to the case of the \emph{Kepler} object of interest KOI-314. This M-dwarf, with a visual magnitude $\sim14$ with a Kepler-band magnitude $K_p = 12.93$, received Keck Observatory HIRES follow-up to confirm the planetary nature of the two transit signals observed in its \emph{Kepler} light curve. We were able to make use of 6 CPS observations to construct a high-resolution spectral template to apply our method. The co-addition procedure, as discussed in Section~\ref{sec:data}, yielded a spectrum with a typical signal-to-noise ratio of $\sim250$ in the red portion, plenty for our purposes. Figure \ref{fig:contours314} shows the results of our analysis with the contours for each pairing in our five parameter fit, $\mall$, $\dvk$, and $d$.

Each parameter is highly correlated, leading to the oblong shaped contours of the figure. This is predominately because the absorption strength of the spectral indices is degenerate in mass and metallicity; the absorption strength can increase with a drop in the effective temperature or an increase in the metal content. Additionally, the shape of the contours between the infrared magnitudes must be consistent with the observed colors ( $J-H, H-K$ etc.).

\begin{figure}[htbp]
   \centering
   \includegraphics[width = .4\textwidth ]{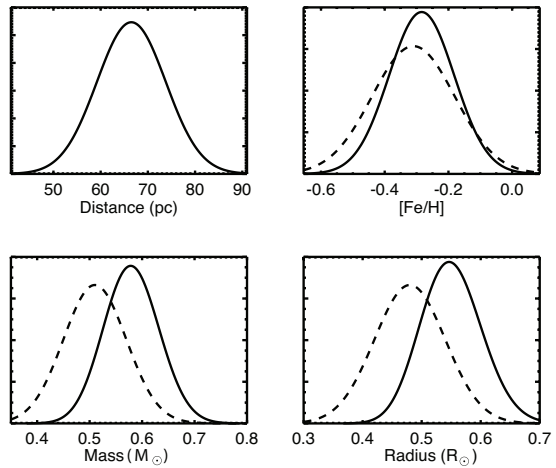}    \caption{Marginalized probability distributions for the properties of KOI=314, using photometric calibrations to convert to mass and [Fe/H]. The \cite{Delfosse2000} is used to convert to mass and the \cite{neves2012} relation is used for metallicity. The distributions from \cite{muirheadkoi} are over plotted with a dashed line, assuming normal distributions with a standard deviation given by the reported uncertainties }
   \label{fig:marg}
\end{figure}

We can then marginalize over the full posterior probability for the five parameters to get probability distributions for the likelihood of each of the parameters. Our estimates are shown in Table~\ref{tab:314} where we have adopted for our uncertainties the scatter we have in reproducing the stellar parameters as demonstrated in Section~\ref{sec:acc}. We have additionally added an empirical estimate of the stellar radius based on the single star mass-radius relation of \cite{boyajian2012} established from interferometric radii measurements of nearby low-mass stars. The table also includes select estimates from \cite{muirheadkoi}, which uses infrared spectra in conjunction with stellar models to derive properties. Additionally, in Figure \ref{fig:marg}, we plot the marginalized distributions for the stellar properties, after converting the photometric properties, $\mk$ and $\dvk$ to the physical properties of mass and [Fe/H]. The top left panel shows our distribution for the distance. The top right panel shows the distribution for [Fe/H], using the calibration of \cite{neves2012} to make the conversion from $\dvk$. The accompanying dashed line is a Gaussian representing the \cite{muirheadkoi} estimate. The bottom left panel shows the distribution for mass, the solid line being our estimate, employing the \cite{Delfosse2000} relation for $\mk$ with the dashed line representing the \cite{muirheadkoi} estimate. Finally, the bottom right panel uses our mass estimate and the mass-radius relation of \cite{boyajian2012} to estimate the radius of the star with the dash-dot line again representing the \cite{muirheadkoi} estimate. Our measurements match those in \cite{muirheadkoi} within the respective uncertainties, giving us confidence in the accuracy of our derived parameters. Combining our estimates with theirs gives a mass estimate of $0.55 \pm 0.04$ $M_{\odot}$, a radius estimate of $0.52 \pm0.04$  $R_{\odot}$ and a metallicity of $-0.29 \pm 0.08$ dex .

\begin{deluxetable}{c c  c }
\tablecaption{ Properties: KOI-314 \label{tab:314}}
\tablehead{ \colhead{ Attribute } &  \colhead{ \cite{muirheadkoi}}  & \colhead{ This Study }  }
\startdata
 $M_{J}$& ... &$ 6.18\pm0.25$ \\
  $M_{H}$&... & $5.57 \pm 0.25$ \\
 $\mk$& ... & $5.39 \pm 0.25$ \\
 $\dvk$& ... & $-0.20 \pm 0.18$ \\
 $d$ & ... & $66.5\pm7.3$ pc \\
 Mass  \tablenotemark{a} & $0.51 \pm 0.06$ $M_{\odot}$  & $0.57 \pm 0.05 $  $M_{\odot}$\\
  Radius\tablenotemark{b} & $0.48 \pm 0.06$ $R_{\odot}$ & $0.54 \pm0.05$  $R_{\odot}$ \\
$[$Fe/H$]  \tablenotemark{c}$ & $-0.31 \pm 0.13$ & $-0.28 \pm 0.10$ 
\enddata
\tablenotetext{a}{ For this work the estimate uses the relations of \cite{Delfosse2000} to convert $\mk$ to mass }
\tablenotetext{b}{ For this work the estimate uses the relations of \cite{boyajian2012} to convert mass to radius }
\tablenotetext{c}{In converting $\dvk$ to [Fe/H], the estimate uses the relation of \cite{neves2012}: [Fe/H] $=0.57\dvk -0.17$. The listed uncertainty does not include the scatter in their relation. }
\end{deluxetable}


\subsection{GJ$\,$3470}

\begin{deluxetable}{c c c}
\tablecaption{ Properties: GJ$\,$3470 System \label{tab:3470}}
\tablehead{ \colhead{ Attribute } & \colhead{ \cite{bonfils12}  }& \colhead{ This Study } }
\startdata
J\tablenotemark{a}& ... & $8.794 \pm 0.019$ \\
H\tablenotemark{a}& ... & $8.206 \pm 0.023$\\
K\tablenotemark{a}& ... &$7.989 \pm 0.023$ \\ 
$a/R_\star$ & $14.9 \pm 1.2$ & ... \\
$\mathrm{Period}$& 3.33714 $\mathrm{days}$ & ...\\
$R_{p}/R_{\star}$& $0.0755 \pm 0.0031$ & ... \\
& &\\
 $d$ & ...  & $29.9\pm_{3.4}^{3.7}$ $\mathrm{pc}$ \\
 Mass   &  $0.54 \pm 0.07 $  $M_{\odot}$ &$0.53 \pm 0.05 $  $M_{\odot}$ \\
  Radius & $0.50 \pm 0.06 $  $R_{\odot}$& $0.50 \pm 0.05 $  $R_{\odot}$\\
$[$Fe/H$]$& ...  & $0.12 \pm 0.12$ 
\enddata
\tablenotetext{a}{ 2MASS photometry of \cite{2MASS}}
\end{deluxetable}

As an additional application of our methods we consider the exoplanet hosting star GJ$\,$3470. In the discovery paper \cite{bonfils12} used HARPS radial velocities and a photometric transit detection to constrain the mass and radius of the planet, adding the system to three other M dwarfs (GL$\,$436, GJ$\,$1214 and KOI-254) with planets that have well-measured mass and radius estimates. The precision of the planet properties for GJ$\,$3470b however was limited by the uncertainty in the stellar properties. Previous studies of GJ$\,$3470 suggested that it is a typical field star on the main sequence, making the methods of this paper applicable \citep{bonfils12}.

In Section~\ref{sec:res} we showed how HARPS archival spectra could be used to measure EWs compatible with the spectral index calibrations of Table \ref{tab:coeffs}. We applied the procedures of Section~\ref{sec:res} on the HARPS spectra of GJ$\,$3470 to obtain a spectrum with a typical signal-to-noise of 72 near the spectral indices. We measured the EWs and were able to independently estimate the distance, mass, [Fe/H] and radius of the star using entirely empirical methods; we used the mass-radius relation of \cite{boyajian2012} to determine the radius from the mass. However, only the first four of the indices of Table \ref{spectable} are useable with the HARPS spectra, reducing the precision of our measurements. Using just the four indices we reproduced the properties of the calibration sample to a RMS of $0.38$ in $\mall$, $0.27$ dex in $\dvk$ and $18\%$ in distance (see methods in Section~\ref{sec:acc}). This RMS corresponds to 0.08 $M_{\odot}$ in mass and 0.15 dex in metallicity using the calibrations of \cite{Delfosse2000} and \cite{neves2012} respectively. To improve the precision of our estimates we look for additional constraints to apply to the stellar properties. 

The broadband photometric methods of \cite{koi254} showed how photometric observations could be combined with transit light curve observables to provide precise estimates of the stellar properties. Following their example we include the reduced semi-major axis of the planet orbit as an additional constraint:

\begin{equation}
\frac{a}{R_{\star}} (M_{\star}, P) = \left(\frac{G}{4\pi^2}\right)^{1/3} \frac{M^{1/3}_{\star}}{R_{\star}(M_{\star})} P^{2/3} \; ,
\label{eq:ar}
\end{equation}

\noindent where the radius is given as a function of mass using the mass-radius relation of \cite{boyajian2012} and we neglect the mass of the planet as very small compared to the stellar mass. The transit light-curve observable, $d_{p}/R_{\star}$, gives the planet-star separation at the time of transit and for the case of a circular orbit matches the reduced semi-major axis of Equation \ref{eq:ar}. We take the period to be well defined as 3.33714 days and the observed $(a/R_{\star})_{o}$ as $14.9 \pm 1.2$ from \cite{bonfils12}. We incorporated this constraint by adding an additional term to the total $\chi^{2}$ of Equation \ref{eq:chitot}, 

\begin{equation}
\chi^{2}_{a} = \frac{  [ (a / R_{\star})_{o}   -  a/R_{\star} (M_{\star}, P)  ]^{2} }{2 \sigma_{ar}^2} \; .
\label{eq:chi_ar}
\end{equation}

\begin{figure}[htbp]
   \centering
   \includegraphics[scale=0.45]{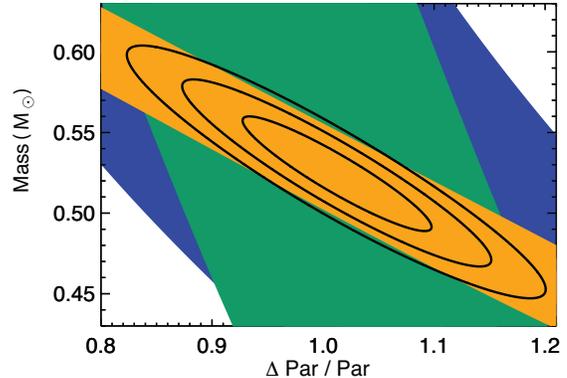} 
   \caption{Contours of the probability distribution for our mass estimate of GJ$\,$3470 as a function of observed parameters with the metallicity and distance fixed to their best estimates, 0.12 dex and 29.9 pc respectively. The outer (blue) region defines the 3$\sigma$ constraint applied by the measurement of $a/R_{\star}$. The middle (green) regions applies the combined 3$\sigma$ constraint of the infrared passbands $JHK$. The thin (orange) region is the combined 3$\sigma$ constraint using our calibration on the measured EWs of the HARPS spectra. The black contours represent the 1$\sigma$, $2\sigma$, and $3\sigma$ contour levels for the entire set of constraints.}
   \label{fig:concord}
\end{figure}

\noindent In implementing the MCMC methods of Section \ref{sec:fit} we use the relations of \cite{Delfosse2000} to go from $\mk$ to $M_{\star}$ in Equation~\ref{eq:chi_ar}. Our estimates, shown in Table \ref{tab:3470}, agree well with the output values derived by the planet-discovery team which combined a transit detection with radial velocity measurements, $0.54 \pm 0.07$ $M_{\odot}$ for the mass  and $0.50 \pm 0.06$ $R_{\odot}$ for the radius, however our stellar properties are more precise. They are also consistent with stellar parameter estimates incorporating a new infrared transit analysis \citep{demory13}. In Figure~\ref{fig:concord} we plot the contours for the total probability distribution in our estimate of the stellar mass as a function of observed quantities at the fixed distance and metallicity of our estimates shown in Table \ref{tab:3470}. The filled overlays show the $3\sigma$ constraints provided by each observation with blue corresponding to $a/R_{\star}$, green to the joint constraint in $JHK$ and orange given by the combined constraints of the EWs.

We can combine our stellar radius measurement with the estimate from \cite{bonfils12} to get a precise stellar radius of $0.50 \pm 0.04$ $R_{\odot}$. Using the transit observable $R_{p} / R_{\star}=0.0755 \pm 0.0031$ from \cite{bonfils12}, we get a planet radius estimate of $R_{p}= 4.12 \pm 0.37$ $R_{\oplus}$, slightly smaller than their estimate of $R_{p}= 4.2\pm 0.6$ $R_{\oplus}$.

\begin{figure}[htbp]
   \centering
   \includegraphics[scale=0.45]{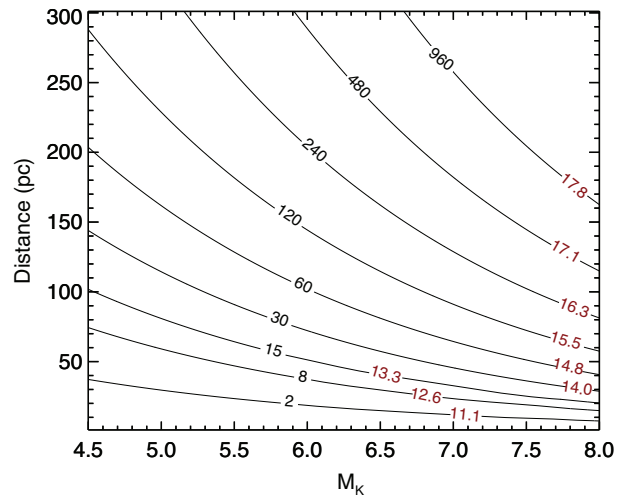} 
   \caption{Contours indicate the approximate minimum total integration time with HIRES, in minutes, for typical low-mass star required to utilize our methods to obtain stellar properties (achieve S/N $\sim70$). The time is shown in the black labels and the corresponding approximate $V$-band magnitude is shown with the red labels}
   \label{fig:obstime}
\end{figure}

\section{Summary and Discussion}

Using high-resolution spectra of nearby M-dwarfs we have developed a new spectroscopic calibration for the physical properties of low-mass stars of types K7-M4. With a high signal-to-noise spectrum and 2MASS photometry we can estimate the mass and the metallicity of the star as well as the distance to the star.

Our methods are based on the integrated spectral flux (EW) at a series of spectral regions sensitive to the known photometric properties of the stars. By measuring the corresponding EWs from the spectra of low-mass stars and making use of the observed infrared magnitudes, we can estimate the photometric properties, $\mall$ and $\dvk$, as well as the distance. Additionally, using known photometric calibrations, these can be converted to mass and metallicity. Our estimates are strongly dependent on the accuracy of the assumed parameters for our calibration stars, however the sample comprises a set of well studied nearby stars with accurate properties. We are thus able to estimate the intrinsic physical properties of low-mass stars without having parallax measurements, independently of stellar models.

We applied our methods to a particular star KOI-314, one of the \emph{Kepler} objects of interest and estimated a mass of $0.58\pm0.05$  $M_{\odot}$, a radius of $0.55\pm^{0.06}_{0.05}$  $R_{\odot}$, a metallicity, [Fe/H], of $-0.28 \pm 0.10$ and a distance of $66.5 \pm 7.3$ pc, where we have adopted for our uncertainties the representative scatter of Section~\ref{sec:acc} and propagated those uncertainties through the empirical calibrations to the uncertainties of the desired physical properties. These estimates are in good agreement with the approach taken by \cite{muirheadkoi}, providing additional observational evidence in corroboration of their model-dependent methods. We were also able to apply our methods to the planet host GJ$\,$3470, making use of archival HARPS spectra and taking advantage of the transit observable $a/R_{\star}$ to narrow in on the stellar properties. We estimated a mass of $0.53 \pm0.05$  $M_{\odot}$, a radius of $0.50\pm0.05$  $R_{\odot}$, a metallicity, [Fe/H], of $0.12 \pm 0.12$ and a distance of $29.9\pm_{3.4}^{3.7}$ pc. These properties are very similar to those determined by \cite{bonfils12} who then estimate an approximate mass of $\sim$14 $M_{\oplus}$ and radius of $\sim$4.2 $R_{\oplus}$ for GJ$\,$3470b.

Although our calibration provides estimates for stellar properties, there are some limitations. We examined the effect of spectral resolution and found that we could use the calibration only for spectral resolutions greater than $\sim$30,000. We also demonstrated how to use the calibration with higher resolution data ($>$50,000) by making use of HARPS archival spectra. Additionally, our calibration sample only spans a particular range of $\mk$ (see Figure \ref{hist}) and should only be used outside that range with caution. This restricts applicability to mostly early type M dwarfs, earlier than about M4, and late K dwarfs. This still spans a fairly broad range of masses from about $\sim 0.7$ to $\sim 0.2$ $M_{\odot}$. 

Our method requires a  high-signal, high-resolution, spectrum of the star. To quantify the necessary signal-to-noise, we used the observations of KOI-314 as a guide in our noise analysis (see Section~\ref{sec:acc}). In Figure \ref{fig:obstime}, we show the approximate minimum total integration time needed, in minutes (on left side), to achieve a signal-to-noise of $\sim70$ and use the techniques presented in this contribution. The corresponding V-band magnitudes are also shown on the contours towards the right side of the plot. As a benchmark, a star with a $V$-band magnitude of 14 would need a total integration time of 30 minutes. It is possible to build up this signal over time by building a composite spectrum. This makes it an ideal method to complement radial velocity surveys of M dwarfs. Many spectra are needed to sample the radial velocities of these stars, so as a byproduct of those observations the physical stellar properties can be determined simultaneously. Additionally, this method can also be applied immediately to archival HIRES data of low-mass stars\footnote{Keck Observatory Archive:\\
http://www2.keck.hawaii.edu/koa/public/koa.php}.

\section*{Appendix}
\section*{Continuum Normalization}

\begin{figure}[htbp]
   \centering
   \includegraphics[scale=0.45]{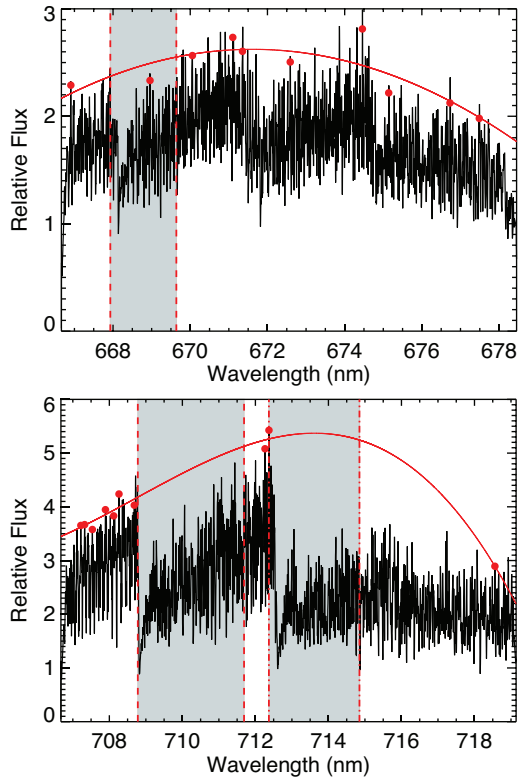}
   \caption{Example spectrum with normalization plotted overtop. Each plot comprises a full order and shows the regions used as spectral indices for computing equivalent widths. Points used for the continuum normalization are shown as filled circles. Chosen points assure that the normalization addresses the convex shape of instrument profile. All of the wiggles are real features in the spectra.}
   \label{continuum}
\end{figure}

The continuum normalization procedure was briefly explained in Section~\ref{speclib} however, we expand upon the details here. The reduced spectra from the HIRES detector include several orders, ten of those corresponding to the red chip of the detector, spanning between 650 nm and 800 nm. Each spectral order is affected by the blaze function of the detector and the overall shape of the stellar spectrum, and we therefore normalize each order separately. For each order, the spectral regions listed in Table \ref{tab:norm}, we first masked out any telluric regions. We then equally divided each region into 10 or 20 (see `Divisions' in Table \ref{tab:norm}) different sections. For each section we ordered the flux and took the top $1\%$ or $2\%$ (see `Percentage' in Table \ref{tab:norm}) level as representative of a pseudo-continuum. We then fit these points with a polynomial of order 2 or 3 (see `Polynomial Order' in Table \ref{tab:norm}) to define our normalization. The number of points in the fit therefore matches the number of bins we used. For each order we divided the spectrum by this polynomial to get the normalized spectrum (see example in Figure \ref{continuum}). Changes in the normalization properties attempt to account for differences in the curvature/symmetry of the spectral continuum/blaze function to achieve an appropriate normalization. To this end the fifth spectral order in Table \ref{tab:norm} does not use the full order in the normalization and instead we applied a narrower range of wavelengths when applying our normalization methods in order to account for the broad and deep absorption bands (see bottom panel of Figure \ref{continuum}). Despite the lack of pseudo-continuum points toward greater wavelengths in the lower panel, the chosen points assure that our normalization accounts for the generally convex shape of the blaze function for each the order. Although there can be issues at the edge of each order, the spectral indices are all located near the central regions so this does not affect the calculated equivalent widths.

\begin{deluxetable}{c c c c}
\tablecaption{ Normalization Properties \label{tab:norm}}
\tablehead{ \colhead{ Region (nm)} & \colhead{ Divisions} & \colhead{ Percentage \% } & \colhead{ Polynomial Order }  }
\startdata
 654.35 - 665.92 & 10 & 1 & 3\\
666.69 - 678.49    & 10 & 1 & 2 \\
679.52 - 691.53  & 10 & 1 & 3 \\
692.85 - 705.08 & 10& 1 & 3\\
706.71 - 719.17  \tablenotemark{a}  & 10 & 1 & 3 \\
721.14 - 733.84 & 10 & 1 &3 \\
736.17 - 749.12& 10 & 1 &3 \\
751.84 - 765.05& 20 & 2 &2 \\
768.19 - 781.67& 10 & 1 &3 \\
785.28 - 796.55& 10 & 1 & 2
\enddata
\tablenotetext{a}{ Within region only the following sections are used for computing normalization: 707.04-708.79 nm, 712.19-712.58 nm, 718.35-718.63 nm }
\end{deluxetable}

We opted not to use continuum regions defined as the linear interpolation of points flanking the absorption region because the spectral regions we identified as sensitive to stellar properties were not always bounded by points of minimal absorption. Our procedure also addresses the difficulties introduced by the shape of the blaze function on the spectrum and by taking a broader region into account for the continuum calculation we can get consistent pseudo-continua across many different spectra. Our application of this procedure to the HARPS data was able to produce equivalent width measurements for several indices in accord with our HIRES measurements, showing how our methods can be used for different data samples (see Section~\ref{sec:res}).

\section*{Acknowledgments}

The authors would like to thank the California Planet Search team for the use of the data, the Keck telescopes support staff for their assistance, Philip Muirhead for his support in this endeavor. The authors would also like to thank the native Hawaiians for the use of their mountain in the advancement of science.

The authors would also like to acknowledge the referee for providing useful comments that strengthened this work.

JSP was supported by a grant from the National Science Foundation Graduate 
Research Fellowship under Grant No. (DGE-1144469).

JAJ acknowledges support from the Sloan Foundation and the David \& Lucile Packard Foundation.

Based on data obtained from the ESO Science Archive Facility under request number(s):
mbottom 48576, jspineda 49793, 49870, 49872, 49874-91, 49893-914, 49916-26.

This publication makes use of data products
from the Two Micron All Sky Survey, which is a
joint project of the University of Massachusetts
and the Infrared Processing and Analysis Cen-
ter/California Institute of Technology, funded by
the National Aeronautics and Space Administra-
tion and the National Science Foundation.

\bibliographystyle{apj}

\bibliography{pbj}

\begin{thebibliography}{42}
\expandafter\ifx\csname natexlab\endcsname\relax\def\natexlab#1{#1}\fi

\bibitem[{{Baraffe} {et~al.}(1998){Baraffe}, {Chabrier}, {Allard}, \&
  {Hauschildt}}]{Baraffe1998}
{Baraffe}, I., {Chabrier}, G., {Allard}, F., \& {Hauschildt}, P.~H. 1998, \aap,
  337, 403

\bibitem[{{Batalha} {et~al.}(2012)}]{Kepler2}
{Batalha}, N.~M., {et~al.} 2012, ArXiv e-prints

\bibitem[{{Bean} {et~al.}(2006){Bean}, {Sneden}, {Hauschildt}, {Johns-Krull},
  \& {Benedict}}]{Bean2006}
{Bean}, J.~L., {Sneden}, C., {Hauschildt}, P.~H., {Johns-Krull}, C.~M., \&
  {Benedict}, G.~F. 2006, \apj, 652, 1604

\bibitem[{{Bonfils} {et~al.}(2005{\natexlab{a}}){Bonfils}, {Forveille},
  {Delfosse}, {Udry}, {Mayor}, {Perrier}, {Bouchy}, {Pepe}, {Queloz}, \&
  {Bertaux}}]{Bonfils2005}
{Bonfils}, X., {Forveille}, T., {Delfosse}, X., {Udry}, S., {Mayor}, M.,
  {Perrier}, C., {Bouchy}, F., {Pepe}, F., {Queloz}, D., \& {Bertaux}, J.-L.
  2005{\natexlab{a}}, \aap, 443, L15

\bibitem[{{Bonfils} {et~al.}(2005{\natexlab{b}}){Bonfils}, {Forveille},
  {Delfosse}, {Udry}, {Mayor}, {Perrier}, {Bouchy}, {Pepe}, {Queloz}, \&
  {Bertaux}}]{BonfilsPlanet}
---. 2005{\natexlab{b}}, \aap, 443, L15

\bibitem[{{Bonfils} {et~al.}(2012)}]{bonfils12}
{Bonfils}, X., {et~al.} 2012, \aap, 546, A27

\bibitem[{{Borucki} {et~al.}(2011)}]{Borucki2011}
{Borucki}, W.~J., {et~al.} 2011, \apj, 736, 19

\bibitem[{{Boyajian} {et~al.}(2012)}]{boyajian2012}
{Boyajian}, T.~S., {et~al.} 2012, \apj, 757, 112

\bibitem[{{Buchhave} {et~al.}(2012)}]{buchhave2012}
{Buchhave}, L.~A., {et~al.} 2012, \nat, 486, 375

\bibitem[{{Butler} {et~al.}(2004){Butler}, {Vogt}, {Marcy}, {Fischer},
  {Wright}, {Henry}, {Laughlin}, \& {Lissauer}}]{Butler2004}
{Butler}, R.~P., {Vogt}, S.~S., {Marcy}, G.~W., {Fischer}, D.~A., {Wright},
  J.~T., {Henry}, G.~W., {Laughlin}, G., \& {Lissauer}, J.~J. 2004, \apj, 617,
  580

\bibitem[{{Chabrier} \& {Baraffe}(2000)}]{Chabrier2000}
{Chabrier}, G., \& {Baraffe}, I. 2000, \araa, 38, 337

\bibitem[{{Coughlin} {et~al.}(2011){Coughlin}, {L{\'o}pez-Morales}, {Harrison},
  {Ule}, \& {Hoffman}}]{coughlin2011}
{Coughlin}, J.~L., {L{\'o}pez-Morales}, M., {Harrison}, T.~E., {Ule}, N., \&
  {Hoffman}, D.~I. 2011, \aj, 141, 78

\bibitem[{{Cutri} {et~al.}(2003)}]{2MASS}
{Cutri}, R.~M., {et~al.} 2003, VizieR Online Data Catalog, 2246, 0

\bibitem[{{Delfosse} {et~al.}(2000){Delfosse}, {Forveille}, {S{\'e}gransan},
  {Beuzit}, {Udry}, {Perrier}, \& {Mayor}}]{Delfosse2000}
{Delfosse}, X., {Forveille}, T., {S{\'e}gransan}, D., {Beuzit}, J.-L., {Udry},
  S., {Perrier}, C., \& {Mayor}, M. 2000, \aap, 364, 217

\bibitem[{{Demory} {et~al.}(2013){Demory}, {Torres}, {Neves}, {Rogers},
  {Gillon}, {Horch}, {Sullivan}, {Bonfils}, {Delfosse}, {Forveille}, {Lovis},
  {Mayor}, {Santos}, {Seager}, {Smalley}, \& {Udry}}]{demory13}
{Demory}, B.-O., {Torres}, G., {Neves}, V., {Rogers}, L., {Gillon}, M.,
  {Horch}, E., {Sullivan}, P., {Bonfils}, X., {Delfosse}, X., {Forveille}, T.,
  {Lovis}, C., {Mayor}, M., {Santos}, N., {Seager}, S., {Smalley}, B., \&
  {Udry}, S. 2013, ArXiv e-prints

\bibitem[{{Hauschildt} {et~al.}(1999){Hauschildt}, {Allard}, \&
  {Baron}}]{Hauschildt1999}
{Hauschildt}, P.~H., {Allard}, F., \& {Baron}, E. 1999, \apj, 512, 377

\bibitem[{{Howard} {et~al.}(2010)}]{Howard2010}
{Howard}, A.~W., {et~al.} 2010, \apj, 721, 1467

\bibitem[{{Johnson} \& {Apps}(2009)}]{JA09}
{Johnson}, J.~A., \& {Apps}, K. 2009, \apj, 699, 933

\bibitem[{{Johnson} {et~al.}(2011){Johnson}, {Apps}, {Gazak}, {Crepp},
  {Crossfield}, {Howard}, {Marcy}, {Morton}, {Chubak}, \& {Isaacson}}]{johnlhs}
{Johnson}, J.~A., {Apps}, K., {Gazak}, J.~Z., {Crepp}, J.~R., {Crossfield},
  I.~J., {Howard}, A.~W., {Marcy}, G.~W., {Morton}, T.~D., {Chubak}, C., \&
  {Isaacson}, H. 2011, \apj, 730, 79

\bibitem[{{Johnson} {et~al.}(2012{\natexlab{a}}){Johnson}, {Gazak}, {Apps},
  {Muirhead}, {Crepp}, {Crossfield}, {Boyajian}, {von Braun}, {Rojas-Ayala},
  {Howard}, {Covey}, {Schlawin}, {Hamren}, {Morton}, {Marcy}, \&
  {Lloyd}}]{koi254}
{Johnson}, J.~A., {Gazak}, J.~Z., {Apps}, K., {Muirhead}, P.~S., {Crepp},
  J.~R., {Crossfield}, I.~J.~M., {Boyajian}, T., {von Braun}, K.,
  {Rojas-Ayala}, B., {Howard}, A.~W., {Covey}, K.~R., {Schlawin}, E., {Hamren},
  K., {Morton}, T.~D., {Marcy}, G.~W., \& {Lloyd}, J.~P. 2012{\natexlab{a}},
  \aj, 143, 111

\bibitem[{{Johnson} {et~al.}(2012{\natexlab{b}})}]{john254}
{Johnson}, J.~A., {et~al.} 2012{\natexlab{b}}, \aj, 143, 111

\bibitem[{{Kirkpatrick} {et~al.}(1991){Kirkpatrick}, {Henry}, \&
  {McCarthy}}]{Kirkpatrick1991}
{Kirkpatrick}, J.~D., {Henry}, T.~J., \& {McCarthy}, Jr., D.~W. 1991, \apjs,
  77, 417

\bibitem[{{L{\'o}pez-Morales} \& {Ribas}(2005)}]{lopezmorales2005}
{L{\'o}pez-Morales}, M., \& {Ribas}, I. 2005, \apj, 631, 1120

\bibitem[{{Mayor} {et~al.}(2003)}]{harps}
{Mayor}, M., {et~al.} 2003, The Messenger, 114, 20

\bibitem[{{Morales} {et~al.}(2010){Morales}, {Gallardo}, {Ribas}, {Jordi},
  {Baraffe}, \& {Chabrier}}]{morales2010}
{Morales}, J.~C., {Gallardo}, J., {Ribas}, I., {Jordi}, C., {Baraffe}, I., \&
  {Chabrier}, G. 2010, \apj, 718, 502

\bibitem[{{Muirhead} {et~al.}(2012{\natexlab{a}}){Muirhead}, {Hamren},
  {Schlawin}, {Rojas-Ayala}, {Covey}, \& {Lloyd}}]{muirheadkoi}
{Muirhead}, P.~S., {Hamren}, K., {Schlawin}, E., {Rojas-Ayala}, B., {Covey},
  K.~R., \& {Lloyd}, J.~P. 2012{\natexlab{a}}, \apjl, 750, L37

\bibitem[{{Muirhead} {et~al.}(2012{\natexlab{b}})}]{koi961}
{Muirhead}, P.~S., {et~al.} 2012{\natexlab{b}}, \apj, 747, 144

\bibitem[{{Neves} {et~al.}(2012){Neves}, {Bonfils}, {Santos}, {Delfosse},
  {Forveille}, {Allard}, {Nat{\'a}rio}, {Fernandes}, \& {Udry}}]{neves2012}
{Neves}, V., {Bonfils}, X., {Santos}, N.~C., {Delfosse}, X., {Forveille}, T.,
  {Allard}, F., {Nat{\'a}rio}, C., {Fernandes}, C.~S., \& {Udry}, S. 2012,
  \aap, 538, A25

\bibitem[{{{\"O}nehag} {et~al.}(2012){{\"O}nehag}, {Heiter}, {Gustafsson},
  {Piskunov}, {Plez}, \& {Reiners}}]{onehag2012}
{{\"O}nehag}, A., {Heiter}, U., {Gustafsson}, B., {Piskunov}, N., {Plez}, B.,
  \& {Reiners}, A. 2012, \aap, 542, A33

\bibitem[{{Perryman} \& others.(1997)}]{Perryman1997}
{Perryman}, M.~A.~C., \& others. 1997, \aap, 323, L49

\bibitem[{{Reid} {et~al.}(1995){Reid}, {Hawley}, \& {Gizis}}]{Reid1995}
{Reid}, I.~N., {Hawley}, S.~L., \& {Gizis}, J.~E. 1995, \aj, 110, 1838

\bibitem[{{Ribas}(2006)}]{ribas2006}
{Ribas}, I. 2006, \apss, 304, 89

\bibitem[{{Rivera} {et~al.}(2005){Rivera}, {Lissauer}, {Butler}, {Marcy},
  {Vogt}, {Fischer}, {Brown}, {Laughlin}, \& {Henry}}]{Rivera2005}
{Rivera}, E.~J., {Lissauer}, J.~J., {Butler}, R.~P., {Marcy}, G.~W., {Vogt},
  S.~S., {Fischer}, D.~A., {Brown}, T.~M., {Laughlin}, G., \& {Henry}, G.~W.
  2005, \apj, 634, 625

\bibitem[{{Rojas-Ayala} {et~al.}(2010){Rojas-Ayala}, {Covey}, {Muirhead}, \&
  {Lloyd}}]{Rojas2010}
{Rojas-Ayala}, B., {Covey}, K.~R., {Muirhead}, P.~S., \& {Lloyd}, J.~P. 2010,
  \apjl, 720, L113

\bibitem[{{Rojas-Ayala} {et~al.}(2012){Rojas-Ayala}, {Covey}, {Muirhead}, \&
  {Lloyd}}]{rojas2012}
---. 2012, \apj, 748, 93

\bibitem[{{Schlaufman} \& {Laughlin}(2010)}]{SS10}
{Schlaufman}, K.~C., \& {Laughlin}, G. 2010, \aap, 519, A105

\bibitem[{{Swift} {et~al.}(2013){Swift}, {Johnson}, {Morton}, {Crepp},
  {Montet}, {Fabrycky}, \& {Muirhead}}]{swift13}
{Swift}, J.~J., {Johnson}, J.~A., {Morton}, T.~D., {Crepp}, J.~R., {Montet},
  B.~T., {Fabrycky}, D.~C., \& {Muirhead}, P.~S. 2013, ArXiv e-prints

\bibitem[{{Terrien} {et~al.}(2012){Terrien}, {Mahadevan}, {Bender},
  {Deshpande}, {Ramsey}, \& {Bochanski}}]{terrien2012}
{Terrien}, R.~C., {Mahadevan}, S., {Bender}, C.~F., {Deshpande}, R., {Ramsey},
  L.~W., \& {Bochanski}, J.~J. 2012, \apjl, 747, L38

\bibitem[{{Tonry} \& {Davis}(1979)}]{Tonry1979}
{Tonry}, J., \& {Davis}, M. 1979, \aj, 84, 1511

\bibitem[{{Voges} {et~al.}(1999){Voges}, {Aschenbach}, {Boller},
  {Br{\"a}uninger}, {Briel}, {Burkert}, {Dennerl}, {Englhauser}, {Gruber},
  {Haberl}, {Hartner}, {Hasinger}, {K{\"u}rster}, {Pfeffermann}, {Pietsch},
  {Predehl}, {Rosso}, {Schmitt}, {Tr{\"u}mper}, \& {Zimmermann}}]{rosat}
{Voges}, W., {Aschenbach}, B., {Boller}, T., {Br{\"a}uninger}, H., {Briel}, U.,
  {Burkert}, W., {Dennerl}, K., {Englhauser}, J., {Gruber}, R., {Haberl}, F.,
  {Hartner}, G., {Hasinger}, G., {K{\"u}rster}, M., {Pfeffermann}, E.,
  {Pietsch}, W., {Predehl}, P., {Rosso}, C., {Schmitt}, J.~H.~M.~M.,
  {Tr{\"u}mper}, J., \& {Zimmermann}, H.~U. 1999, \aap, 349, 389

\bibitem[{{Vogt} {et~al.}(1994)}]{vogt1994}
{Vogt}, S.~S., {et~al.} 1994, in Society of Photo-Optical Instrumentation
  Engineers (SPIE) Conference Series, Vol. 2198, Society of Photo-Optical
  Instrumentation Engineers (SPIE) Conference Series, ed. D.~L. {Crawford} \&
  E.~R. {Craine}, 362

\bibitem[{{Woolf} \& {Wallerstein}(2005)}]{Woolf2005}
{Woolf}, V.~M., \& {Wallerstein}, G. 2005, \mnras, 356, 963

\end{thebibliography}

\end{document}